\RequirePackage[l2tabu,orthodox]{nag}
\documentclass[12pt,letterpaper]{article}
\usepackage[T1]{fontenc}
\usepackage[utf8]{inputenc}
\usepackage[strict,autostyle]{csquotes}
\usepackage[USenglish]{babel}
\usepackage{microtype}
\usepackage{crimson}
\usepackage{authblk}
\usepackage{booktabs}
\usepackage{endnotes}
\usepackage{graphicx}
\usepackage{hyperref}
\usepackage{natbib}
\usepackage{url}

\graphicspath{{./fig/}}
\hypersetup{
	pdfauthor={Geoff Boeing},
	pdftitle={Estimating Local Daytime Population Density from Census and Payroll Data},
	pdfsubject={Estimating Local Daytime Population Density from Census and Payroll Data},
	pdfkeywords={GIS, Spatial Analysis, Population, Density, Data Science, Visualization, LEHD, LODES, Census},
	pdffitwindow=true,
	breaklinks=true,
	colorlinks=true,
	citecolor=black,
	linkcolor=black,
	filecolor=black,      
	urlcolor=blue}

\begin{document}
	
\title{Estimating Local Daytime Population Density from Census and Payroll Data\footnote{Citation info: Boeing, G. 2018. \enquote{Estimating Local Daytime Population Density from Census and Payroll Data}. \textit{Regional Studies, Regional Science}, in press. \href{https://doi.org/10.1080/21681376.2018.1455535}{doi:10.1080/21681376.2018.1455535}}}
\date{May 2018}
\author[]{Geoff Boeing}
\affil[]{University of California, Berkeley}

\maketitle

\begin{abstract}
Daytime population density reflects where people commute and spend their waking hours. It carries significant weight as urban planners and engineers site transportation infrastructure and utilities, plan for disaster recovery, and assess urban vitality. Various methods with various drawbacks exist to estimate daytime population density across a metropolitan area, such as using census data, travel diaries, GPS traces, or publicly available payroll data. This study estimates the San Francisco Bay Area’s tract-level daytime population density from US Census and LEHD LODES data. Estimated daytime densities are substantially more concentrated than corresponding nighttime population densities, reflecting regional land use patterns. We conclude with a discussion of biases, limitations, and implications of this methodology.
\end{abstract}

When we study urban density, we often mean nighttime population density---where people live and sleep. However, urban planners and engineers are equally interested in daytime density---where people commute and spend their waking hours---to site transportation infrastructure and utilities, plan for disaster recovery, and assess urban vitality \citep{schmitt_estimating_1956}. Planners might estimate local daytime population density across a metropolitan area using, for example, American Community Survey (ACS) data, travel diaries, or publicly-available payroll data. This study estimates the San Francisco Bay Area's tract-level daytime population density from US census and payroll data then explores biases, limitations, and implications. This methodology easily scales nationwide.

We use three input data products: the 2010 US census TIGER/Line tracts shapefile with DP1 attributes\endnote{US census tracts 2010 TIGER/Line shapefile with DP1, available at \url{http://www2.census.gov/geo/tiger/TIGER2010DP1/Tract_2010Census_DP1.zip}}, the US census bureau's 2010 states shapefile\endnote{Census bureau 2010 US states 1:500,000 resolution shapefile, available at \url{http://www2.census.gov/geo/tiger/GENZ2010/gz_2010_us_040_00_500k.zip}}, and the 2010 Longitudinal Employer-Household Dynamics Origin-Destination Employment Statistics\endnote{California LEHD LODES7 2010 origin-destination main data, available at \url{https://lehd.ces.census.gov/data/lodes/LODES7/ca/od/ca_od_main_JT00_2010.csv.gz}} (LODES) for California. LODES is an administrative payroll enumeration of jobs with both workplaces and residences (geocoded at the block level) in the state. However, if the employer has multiple workplaces, the reported payroll-based workplace may not be the one to which the employee actually commutes \citep{nelson_economic_2016}.

We prefer the 2010 demographic data to more recent ACS data because the latter's tract-level estimates encompass five-year rolling averages. Accordingly we prefer not to compare 2014 LODES data to 2010-2014 ACS data as the Bay Area experienced substantial housing, economic, and demographic upheaval over this timeframe, patterns obscured in the ACS rolling averages \citep{boeing_new_2017}. To avoid inconsistent comparison, we opt for more stale---but more accurate and comparable---decennial data \citep{macdonald_american_2006,spielman_patterns_2014}.
 
LODES is notoriously noisy (and synthetic) so we aggregate and sum the origin-destination pairs to the tract level, at which it converges reasonably well to the observed distribution \citep{spear_improving_2011}. Then we merge\endnote{We use Python and a Jupyter notebook to conduct this analysis and produce the choropleth map. Code available on GitHub at \url{https://github.com/gboeing/data-visualization/tree/master/daytime-population-density}} these data with Bay Area tract-level population, and calculate daytime population density $D$ for each tract $t$ as:

\begin{equation}
	\label{eq:density}
	D_t = \frac{P_t + I_t - O_t}{A_t}
\end{equation}

\begin{figure}[htbp]
	\includegraphics[width=\textwidth]{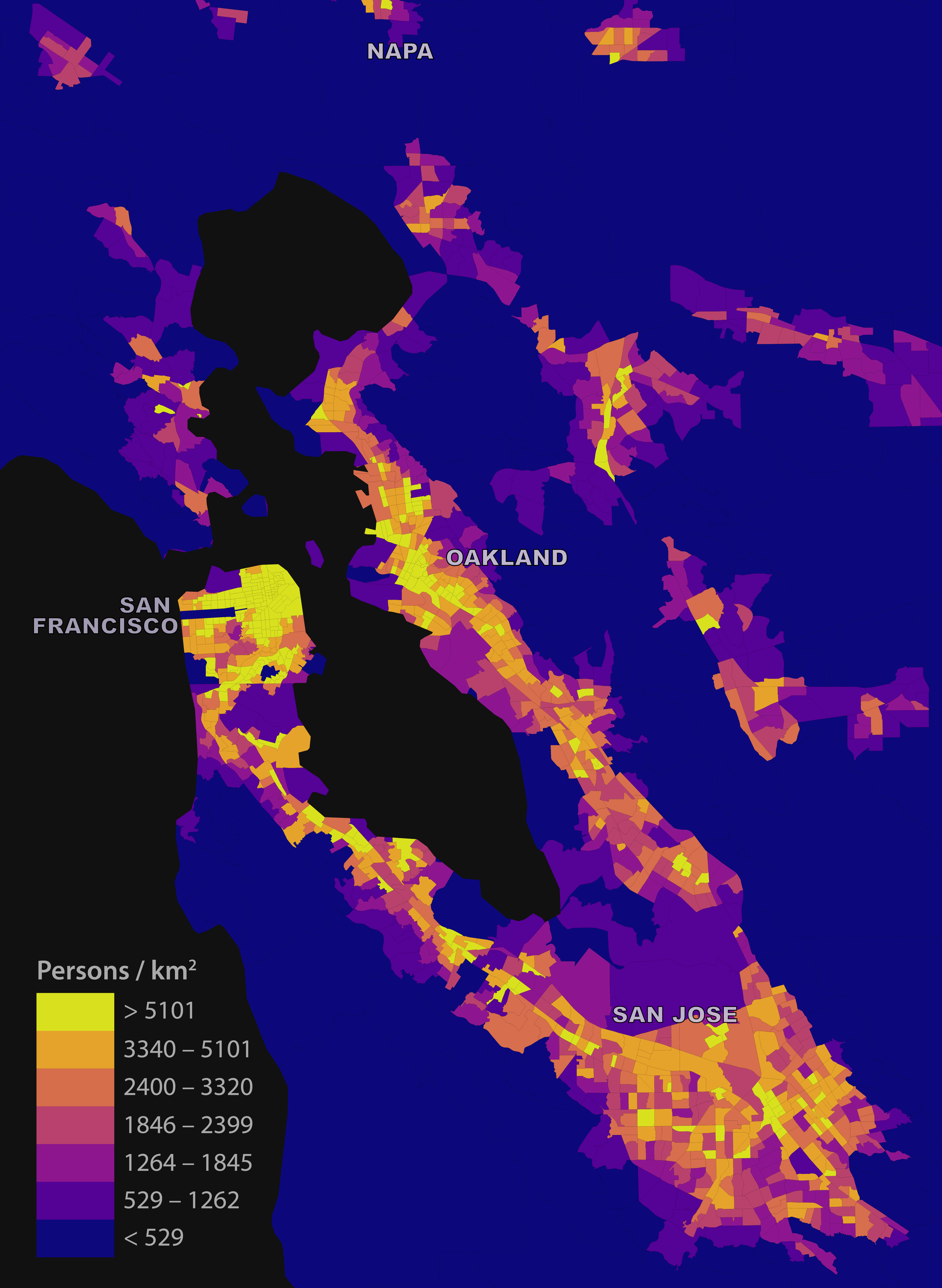}
	\caption{Estimated daytime population density in the San Francisco Bay Area.}
	\label{fig:map}
\end{figure}

Where $P_t$ is the tract's population, $I_t$ is its inbound commuters, $O_t$ is its outbound commuters, and $A_t$ is its land area (km\textsuperscript{2}). We map these tracts in Figure \ref{fig:map} by trimming their geometries to California's state shapefile extents to make the bay legible (census tracts otherwise cover it) and because we normalized by land area. This does however raise an interesting question about the large population of houseboats off the shores of Sausalito. Finally, we produce an interactive web map available online\endnote{See \url{http://geoffboeing.com/2017/12/estimating-daytime-population-density/}}.

\begin{table}[tbp]
	\centering
	\caption{Census tracts with highest daytime population densities (persons/km\textsuperscript{2}) in the San Francisco Bay Area.}
	\label{tab:highest_densities}
	\begin{tabular}{lrrrr}
		\toprule
		Tract &  Population &  Daytime Pop &  Land Area (km\textsuperscript{2}) &  Daytime Density \\
		\midrule
		06075011700 &             1783 &        70728 &      0.556 &       127198 \\
		06075020100 &             6172 &        42635 &      0.446 &        95652 \\
		06075012301 &             2734 &         8006 &      0.092 &        86882 \\
		06075011800 &             1500 &         4850 &      0.057 &        85743 \\
		06075061100 &             4307 &        19051 &      0.240 &        79424 \\
		06075030101 &             4233 &        22416 &      0.346 &        64768 \\
		06075017801 &             3499 &        13051 &      0.214 &        60971 \\
		06075061500 &            11502 &        92865 &      1.670 &        55617 \\
		06075012502 &             3821 &         3319 &      0.061 &        54162 \\
		06075012302 &             3073 &         4829 &      0.093 &        52122 \\
		\bottomrule
	\end{tabular}
\end{table}

The median daytime population density across all Bay Area tracts is 2,097 persons/km\textsuperscript{2} but the distribution has an extreme right tail: the standard deviation $\sigma$ of Figure \ref{fig:map}'s highest quantile (15,330) far exceeds the average $\sigma$ across its other quantiles (249).
Table \ref{tab:highest_densities} lists the 10 tracts with the highest daytime densities, all of which are within the city of San Francisco. The densest tract---comprising the central Financial District and Union Square neighborhoods---contains over 127,000 persons/km\textsuperscript{2} during the day, when its population swells by a factor of 40. Among these 10 tracts, only one has a net outflow of commuters. Region-wide, tract daytime population's Gini coefficient is 70\% higher than that of nighttime population (0.36 vs 0.21), suggesting that people concentrate into fewer tracts during the day, but disperse more evenly among all tracts when they return home at night.

We cannot calculate confidence intervals to assess our estimates in a meaningful way from these data, as they are not sampled. The decennial census is a complete enumeration and the LODES data is an administrative payroll enumeration. Had we used ACS data, we could have looked at sample estimates and standard errors, but this still would not account for the LODES enumeration. More importantly, we systematically ignore or miscount the flow of tourists, shoppers, students, telecommuters, the self-employed, government workers, and populations less legible to these data products, such as certain minority groups and the homeless \citep{spear_improving_2011}. According to its post-enumeration survey, the 2010 census systematically overcounted white Americans and undercounted black and Hispanic Americans as well as renters \citep{groves_how_2012}.

Nevertheless, Figure \ref{fig:map}'s density patterns conform to expectations. The Bay Area's polycentric urban cores clearly stand out, but there are anomalies. Due to its student and government worker populations (which LODES ignores), UC Berkeley's campus shows an absurdly low daytime density. What about other places that would be prime locations for urban vitality, but whose daytime populations are drastically underrepresented by residence and commute, such as public plazas, parks, and high schools? Alternative data, such as mobile phone traces, could tell other sides of this story, but are biased toward certain populations and can be difficult to acquire. Finally, not all urban spaces are created equal: the characteristics, culture, and type of density matter. An office building and a public square could exhibit similar daytime density while contributing very differently to urban vitality, let alone posing different problems for infrastructure engineering and evacuation planning.

Human density plays a recognized role in city vitality, reduced energy consumption and greenhouse gas emissions, and increased pooling and matching agglomeration efficiencies. This study discussed one method of estimating daytime density from census population data and LODES payroll data, producing a rough estimate biased toward commuters and against less-legible daily population flows.

\theendnotes

\bibliographystyle{apalike}
\bibliography{rsrs-daytime-density}

\begin{thebibliography}{}

\bibitem[Boeing and Waddell, 2017]{boeing_new_2017}
Boeing, G. and Waddell, P. (2017).
\newblock New {Insights} into {Rental} {Housing} {Markets} across the {United}
  {States}: {Web} {Scraping} and {Analyzing} {Craigslist} {Rental} {Listings}.
\newblock {\em Journal of Planning Education and Research}, 37(4):457--476.

\bibitem[Groves, 2012]{groves_how_2012}
Groves, R. (2012).
\newblock How {Good} was the 2010 {Census}? {A} {View} from the
  {Post}-{Enumeration} {Survey}.\\
\newblock
  https://www.census.gov/newsroom/blogs/director/2012/05/how-good-was-the-2010-census-a-view-from-the-post-enumeration-survey.html.

\bibitem[Macdonald, 2006]{macdonald_american_2006}
Macdonald, H. (2006).
\newblock The {American} {Community} {Survey}: {Warmer} ({More} {Current}), but
  {Fuzzier} ({Less} {Precise}) than the {Decennial} {Census}.
\newblock {\em Journal of the American Planning Association}, 72(4):491--503.

\bibitem[Nelson and Rae, 2016]{nelson_economic_2016}
Nelson, G. and Rae, A. (2016).
\newblock An {Economic} {Geography} of the {United} {States}: {From} {Commutes}
  to {Megaregions}.
\newblock {\em PLoS ONE}, 11(11):e0166083.

\bibitem[Schmitt, 1956]{schmitt_estimating_1956}
Schmitt, R. (1956).
\newblock Estimating {Daytime} {Populations}.
\newblock {\em Journal of the American Institute of Planners}, 22(2):83--85.

\bibitem[Spear, 2011]{spear_improving_2011}
Spear, B. (2011).
\newblock Improving {Employment} {Data} for {Transportation} {Planning}.
\newblock Technical Report NCHRP 08-36 Task 098, Cambridge Systematics,
  Cambridge, MA.

\bibitem[Spielman et~al., 2014]{spielman_patterns_2014}
Spielman, S., Folch, D., and Nagle, N. (2014).
\newblock Patterns and causes of uncertainty in the {American} {Community}
  {Survey}.
\newblock {\em Applied Geography}, 46:147--157.

\end{thebibliography}

\end{document}